\RequirePackage{fix-cm}
\documentclass[pdftex,twocolumn,epjc3]{preprint}   
\smartqed  
\RequirePackage{graphicx}
\RequirePackage{amsmath, amssymb}
\RequirePackage{amsfonts}
\RequirePackage{lipsum}
\begin{document}

\title{Limits on Kaluza-Klein dark matter annihilation in the Sun from recent IceCube results
}

\author{M. Colom i Bernadich\thanksref{addr1}
        \and
        C. P\'erez de los Heros\thanksref{e2,addr1}
}

\thankstext{e2}{e-mail: cph@physics.uu.se}

\institute{Department of Physics and Astronomy,  Uppsala University.\\
  Box 516, SE-751 20 Uppsala, Sweden. \label{addr1}
}

\maketitle

\begin{abstract}
  We interpret recent IceCube results on sear\-ches for dark matter accumulated in the sun in terms of the lightest  
  Kaluza-Klein excitation (assumed here to be the Kaluza-Klein photon, $B^1$), obtaining improved limits on the annihilation
  rate in the Sun, the resulting neutrino flux at the Earth and on the  $B^1$-proton cross-sections, for $B^1$ masses in the range 
  30--3000 GeV. These results improve previous results from IceCube in its 22-string configuration by up to an 
  order of magnitude, depending on mass, but also extend the results to $B^1$ masses as low as 30 GeV. 
  
\end{abstract}

\section{Introduction}\label{intro}
There are many astrophysical and cosmological observations that point to the existence of a dark matter component as a key
constituent of the Universe. Constraints on the amount of baryons in the Universe from CMB measurements, from 
measurements of the abundance of primordial light elements and from searches for dark objects using
microlensing have practically ruled out the possibility that dark matter consists of known Standard Model particles~\cite{Garrett:2010hd}.
In the most popular picture, dark matter is composed by non-relativistic Weakly Interacting Massive Particles (WIMPs)
of yet unknown nature~\cite{Bertone:2004pz}.  Among the many theories beyond the Standard Model 
of particle physics that predict new particles that could be viable dark matter candidates, Kaluza-Klein type models~\cite{Kaluza:1921tu,Klein:1926tv}
with universal extra dimensions (UED)\footnote{Universal in this context meaning that all the Standard Model fields are free to propagate
 also in the new dimension.} provide a WIMP in the  Kaluza-Klein photon ($B^1$), the first excitation of the scalar gauge boson
in the theory~\cite{Cheng:2002ej,Hooper:2002gs,Servant:2002aq,Hooper:2007qk,Blennow:2009ag,Flacke:2017xsv}. Usually denoted as the lightest Kaluza-Klein
particle (LKP), it can have a mass in the range from a few hundred GeV (limit from rare decay processes~\cite{Freitas:2008vh,Haisch:2007vb}) to a few
TeV (to avoid overclosing the Universe), the mass being proportional to 1/R, where R is the size of the extra dimension.

LKPs in the galactic halo will suffer the same fate as any other WIMP. Assuming a non-zero  $B^1$--proton scattering
cross section, $B^1$'s in the galactic halo with Sun-crossing orbits can loose energy through interaction with the matter in the
Sun and eventually sink into its core, where they would accumulate and annihilate into Standard Model particles~\cite{Spergel:1984re,Press:1985ug,Gaisser:1986ha}, which in turn can lead to
a detectable neutrino flux. Neutrino telescopes like AMANDA, IceCube, ANTARES and Baikal have searched for signatures of dark matter in this way, mainly
focusing on the SUSY neutralino as WIMP candidate~\cite{IceCube:2011aj,Aartsen:2016zhm,Adrian-Martinez:2016gti,Avrorin:2014swy}. Additionally, both IceCube and ANTARES 
have set limits to the spin-dependent LKP-proton cross section~\cite{Abbasi:2009vg,Zornoza:2013ema}. The IceCube limits are based on the event selection and
analysis searching for WIMP dark matter performed in~\cite{Abbasi:2009uz}. 

 In this letter we use the latest dark  matter searches from the Sun by IceCube~\cite{Aartsen:2016zhm} to improve the limits on the
 Kaluza-Klein photon cross section with protons. Given  the increase in detector size (86 strings versus 22 in the
 previous IceCube analysis) and lifetime (104~d versus 532~d) the results presented in this letter improve those in~\cite{Abbasi:2009vg} by  
 up to an order of magnitude. Additionally, the presence of the low-energy subdetector DeepCore allows to lower the explored LKP mass down to 30 GeV, compared
 to 250~GeV, the lowest mass studied in~\cite{Abbasi:2009vg}.

\section{LKP signatures from the Sun}\label{sec:signal}
 
\begin{table}[t!]
  \begin{center}
  \caption{Table with the branching ratios of the main annihilation channels of $B^1B^1\rightarrow X\overline{X}$ in terms of two values of the
    quark splitting mass. The last channel corresponds to a Higgs-anti-Higgs pair. Source:~\cite{Hooper:2002gs}}
\label{tab:BRs}
\begin{tabular}{ ccc } 
\hline\noalign{\smallskip}
Channel & $\Delta m=0$ & $\Delta m=0.14$\\
\noalign{\smallskip}\hline\noalign{\smallskip}
$\nu_{e}\overline{\nu}_e,\nu_{\mu}\overline{\nu}_\mu,\nu_{\tau}\overline{\nu}_\tau$ & 0.012 & 0.014 \\
$e^+e^-,\mu^+\mu^-,\tau^+\tau^-$ & 0.20 & 0.23 \\
$u\overline{u},c\overline{c},t\overline{t}$ & 0.11 & 0.077 \\
$d\overline{d},s\overline{s},b\overline{b}$ & 0.007 & 0.005 \\
$\phi\overline{\phi}$ & 0.023 & 0.027 \\
\noalign{\smallskip}\hline
\end{tabular}
\end{center}
\end{table}

Table~\ref{tab:BRs} shows the theoretical branching ratios of the $B^1$ self-annihilation processes in terms of the quark splitting mass:
\begin{equation}
  \Delta m=\frac{m_{q^1}-m_{B^1}}{m_{B^1}},
\end{equation}
where $m_{q^1}$ is the first fermion excitation in the Kaluza-Klein theory within Universal Extra Dimensions~\cite{Hooper:2002gs}.
The mass at which the LKP can be a good dark matter candidate depends on this splitting, which drives the co-annihilation with the higher KK modes which, 
in turn, determines the relic abundance~\cite{Belanger:2010yx}. Among the final
annihilation products, only the weakly interacting neutrinos are able to escape from the Sun and be detected in neutrino observatories on Earth.
Neutrinos resulting from these annihilations are also expected to be easily distinguishable from thermonuclear reaction products because the value
of $B^1$ mass needed for it to be a good dark matter candidate 
($m_{B^1}> {\cal{O}} 100$ GeV)\footnote{There are experimental limits on the lowest allowed mass for $B^1$ that we discuss below.}.

Assuming that the capture and annihilation rates of $B^1$ particles in the Sun, $\Gamma_C$ and $\Gamma_A$, have reached an equilibrium~\cite{Hooper:2002gs,Wikstrom:2009kw}, the relationship between them can be written as $\Gamma_A=\frac{1}{2}\Gamma_C$, 
where the one half factor comes from the fact that one annihilation requires two captured $B^1$. This means that, assuming a particular
velocity distribution for
dark matter in the solar neighborhood and a solar structure model, the $B^1$--proton scattering cross section, which drives the capture, is proportional to the
annihilation rate,
\begin{equation}
  \label{eq:Xsection}
  \sigma^{i}=\lambda^{i}\left(m_{B^1}\right)\Gamma_A
\end{equation}
where the proportionality constant $\lambda^{i}\left(m_{B^1}\right)$ depends on the mass of the $B^1$, and the superscript $i$ can take the values SD (for spin-dependent cross section) or SI
(for spin-independent cross section). The neutrinos produced in the core of the Sun have to be numerically propagated to a detector on Earth to predict the detectable
neutrino flux per unit area and time in the detector, $\Phi_\nu$, which is proportional to the annihilation rate,
$\Phi_{\nu}=\eta \left(m_{B^1}\right)\Gamma_A$. Such neutrino propagation must also take into account the solar composition,
  neutrino interaction processes with matter (such as absorption, re-emission, decays of secondary particles into neutrinos, etc.) and neutrino oscillations.

\begin{figure}[!t]
\includegraphics[width=\columnwidth]{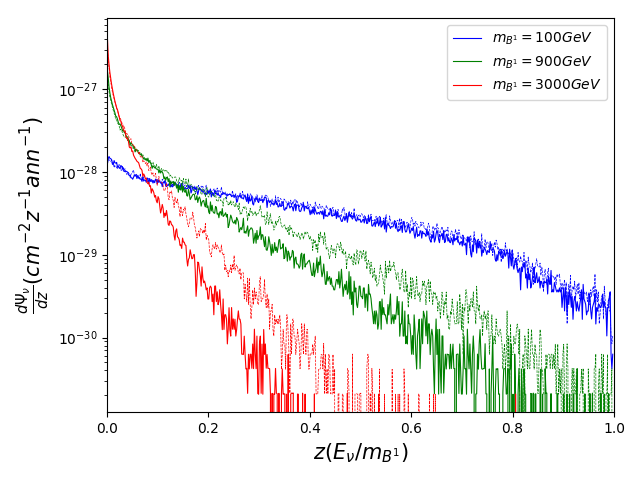}
\caption{Predicted energy spectra of muon neutrinos (solid lines) and anti-muon neutrinos (dashed lines) at the detector from ${B^1}$ annihilations, for three different values of $m_{B^1}$. An extra peak at the end of the spectra produced by neutrinos that escape the Sun without interacting has been omitted from the plot for the sake of legibility, but it is considered in the calculations. Note the normalization in relative energy units.}
\label{fig:spectrum}
\end{figure}

We simulated one million annihilation events at the core of the Sun and we propagated the neutrinos to a detector in the ice at latitude $90^\circ$ S during the austral winter using {\texttt{WimpSim}}~\cite{Edsjo:XXX}, for different assumed values of $m_{B^1}$.  The simulations themselves do not include any information on the capture conditions or $\Gamma_A$, but they do require a solar structure model as an input for the propagation of neutrinos, in this case the one from \cite{Serenelli:2009yc}. Figure~\ref{fig:spectrum} shows the energy distribution of the muon (plus anti-muon) neutrino fluxes at the Earth per unit area $A$ and per annihilation in the Sun, as a function
of reduced energy $z=E_\nu/m_{B^1}$,
\begin{equation}
  \label{eq:diffflux}
  \frac{d\Psi_\nu}{dz}=\frac{dN_\nu}{dAdN_Adz}
\end{equation}
for some values of $m_{B^1}$.
As expected, heavier $B^1$ particles produce steeper profiles in energy because of absorption of high-energy neutrinos on their way out of the Sun. The convolution of those spectra
with the effective area of the detector gives a prediction of the number of events $\mu_s$ per number of annihilations $N_A$, expected at the detector
\begin{equation}
  \label{eq:n_events}
  \frac{d\mu_s}{dN_A}=\int_0^1\frac{d\Psi_\nu}{dz}A_{eff}\left(z\right)dz
\end{equation}

\begin{figure}[t!]
\centering
\includegraphics[width=\columnwidth]{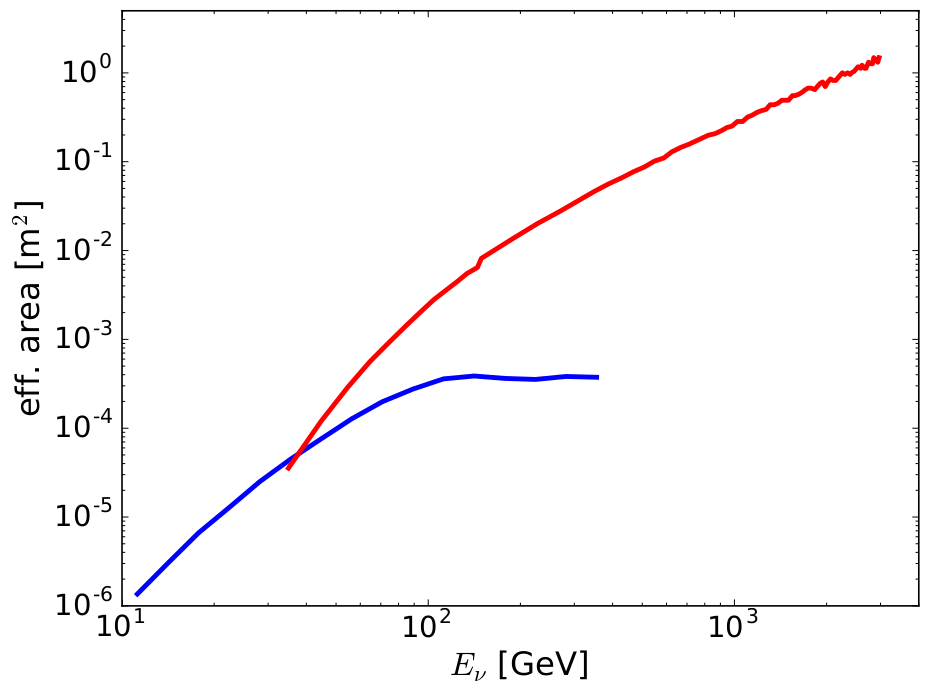}
\caption{The combined muon and anti-muon neutrino effective area for both the DeepCore (blue) and main IceCube array (red) for horizontally arriving neutrinos
  as a function of the neutrino energy. Uncertainties are not shown, but they can be as large as $30\%$. Source:~\cite{Aartsen:2016zhm}}
\label{fig:aeff}
\end{figure}

\begin{figure}[t!]
\centering
\includegraphics[width=\columnwidth]{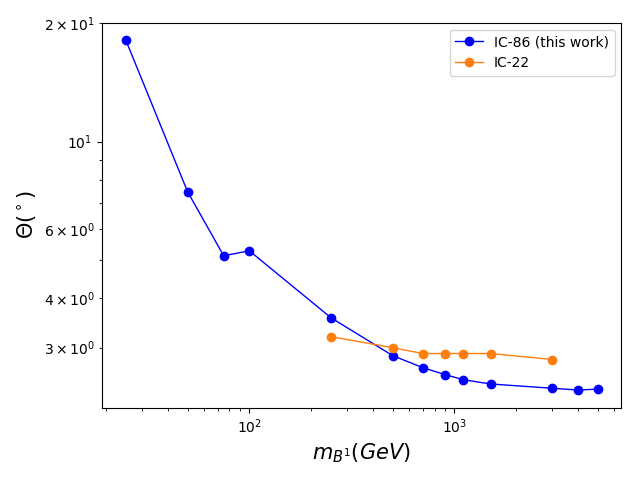}
\caption{Median angular resolution as a function of $B^1$ mass for the 86-string configuration (blue) and the 22-string configuration (orange). The bump at $m_{B^1}=100$ GeV in the former is due to the transition from DeepCore resolution to IceCube resolution~\cite{Aartsen:2016zhm}}
\label{fig:ang_res}
\end{figure}

In order to take into account the finite angular resolution of the detector,  the direction of each event has been randomly smeared using the median
angular resolution of IceCube shown in Figure~\ref{fig:ang_res}, which has been computed from the median muon neutrino energy at every
$m_{B^1}$ and the corresponding mean angular separation between the original neutrino and the reconstructed muon trajectories (see~\cite{Aartsen:2016zhm} for
the explicit relation). $\Theta$ is used to spread the predicted signal (\ref{eq:n_events}) with a 2D Gaussian distribution centered around the Sun on the celestial sphere:
\begin{equation}
  \label{eq:distribution}
  \frac{d\mu_s}{dN_Ad\psi}=\frac{d\mu_s}{dN_A}\frac{C}{\Theta^2}e^{-\psi^2/2\Theta^2}\sin{\psi}
\end{equation}
where $\psi$ is the angle from the Sun ($\psi_\odot=0$), and $C$ a normalization constant. The expected angular distribution of the signal, for a few ${B^1}$ masses as a function of angle with respect to the Sun position is shown in figure~\ref{fig:data}.

\begin{figure}[t!]
\centering
\includegraphics[width=\columnwidth]{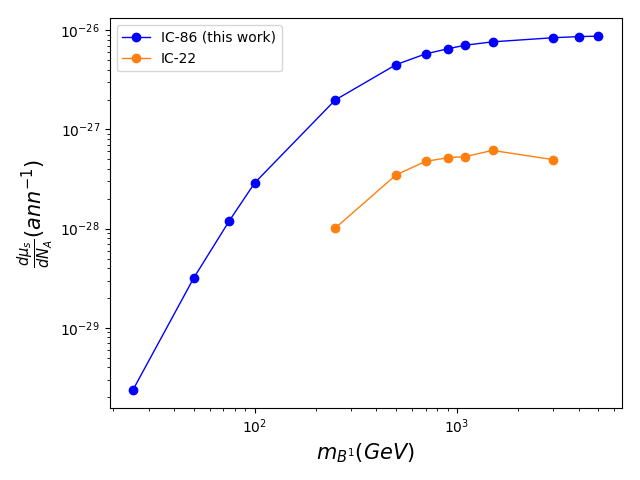}
\caption{Expected number of events in IceCube per annihilation as a function of $B^1$ mass.
  The blue dots show the result of this work (using equation~\ref{eq:n_events} with the effective area of reference~\cite{Aartsen:2016zhm}),
  while the orange dots show, for comparison, the result using the effective area of the 22-string IceCube configuration~\cite{Abbasi:2009uz},
  which was the basis for the previous Kaluza-Klein analysis~\cite{Abbasi:2009vg}. Lines are just a linear interpolation between the dots to guide the eye.}
\label{fig:n_events}
\end{figure}

\section{Data selection}

We use the muon-neutrino effective area corresponding to the latest solar dark matter search from IceCube~\cite{Aartsen:2016zhm}, shown in Figure~\ref{fig:aeff}.
  The figure shows the effective areas for IceCube and DeepCore for muon neutrinos as a function of neutrino energy, $E_\nu$, for the case of 
  neutrinos arriving near the horizon, since the Sun is always close to the horizon in the South Pole. For this analysis, both curves are summed and the
  entire detector is treated as one single array. Only muon events are considered here since the long muon tracks allow pointing back to
  the Sun. \textit{Muon} in what follows refer to both muons and anti-muons since IceCube can not distinguish between particles and antiparticles.

\begin{figure}[t!]
\centering
\includegraphics[width=\columnwidth]{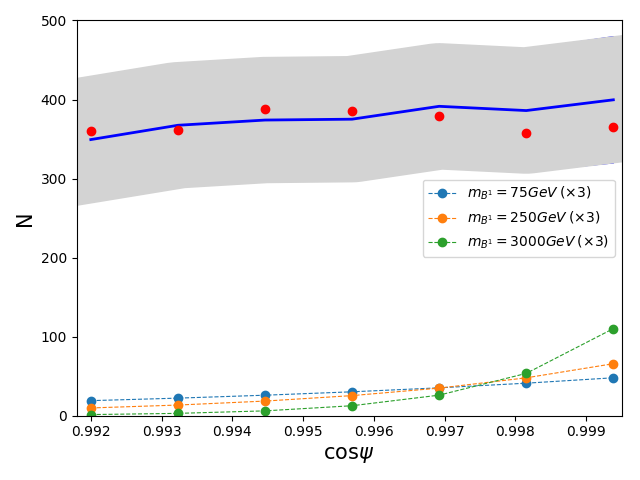}
\caption{Expected background (dark blue line) within the range of maximum systematic uncertainties (shaded area), the actual number
  of detected events at every angular bin (red dots) and the expected signal (multiplied by three for visualization) for some values of $m_{B^1}$
  (light blue, orange and green), as a function of angle with respect to the Sun. Data from~\cite{Aartsen:2016zhm}.}
\label{fig:data}
\end{figure}

The predicted signal (equation~\ref{eq:n_events}) can be compared with the actual observed signal rate to compute $\Gamma_A$:
\begin{equation}
  \frac{d\mu_s}{dt}=\frac{dN_A}{dt}\frac{d\mu_s}{dN_A}=\Gamma_A\frac{d\mu_s}{dN_A}
  \label{eq:mu_s}
\end{equation}
and $\Gamma_A$ is used to calculate the $B^1$--proton scattering cross sections through equation~(\ref{eq:Xsection}).
The conversion factors $\lambda^{i}\left(m_{B^1}\right)$ are obtained with the {\texttt{DarkSUSY}} software \cite{Bringmann:2018lay},
and include the information on the solar structure and the dark matter velocity distribution in the solar neighborhood~\cite{Wikstrom:2009kw}.
Figure~\ref{fig:n_events} shows the expected number of events at the detector per annihilation as a function of $B^1$ mass calculated
with equation~(\ref{eq:n_events})

We use the data from Figure~6 in~\cite{Aartsen:2016zhm}, which we present here combined for DeepCore and IceCube in Figure~\ref{fig:data},  
in order to extract limits on the spin-dependent $B^1$-proton cross section as a function of $B^1$ mass. The data set was obtained during 532 days 
of exposure between May 2001 and May 2014. As thoroughly explained in~\cite{Aartsen:2016zhm}, several filters were applied to the data sample in order 
to minimize the presence of background. Data was only taken into account if measured during austral winters, when the Sun is below the horizon, in order 
to avoid overlap with \textit{atmospheric muons} originating from cosmic-ray induced showers in the atmosphere above the detector. For this reason, only muons with upward trajectories 
were selected. Still, \textit{atmospheric muon neutrinos} created at any declination can cross the Earth without interacting and reach the detector from 
below, being an irreducible background. Boosted Decision Trees were used to maximize signal separation and reduce background.

\begin{figure}[t]
\centering
\includegraphics[width=\columnwidth]{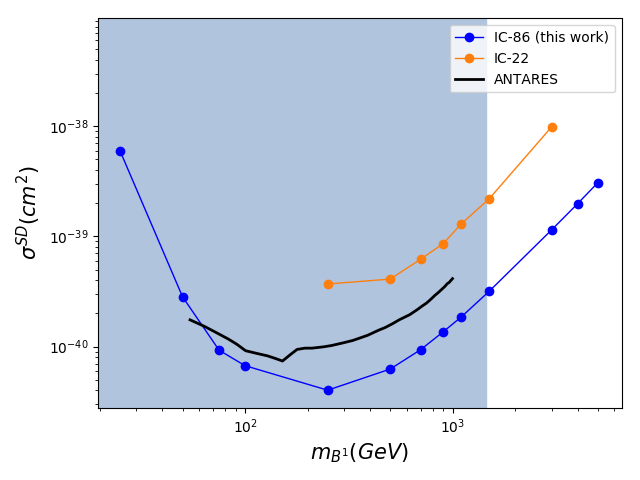}
\caption{90\% confidence level upper limit on the spin-dependent $B^1$--proton scattering cross section  as a function of $B^1$ mass.
  Blue: this work. Orange: previous IceCube limit on LKP cross section from~\cite{Abbasi:2009vg}. Black line: limits from ANTARES~\cite{Zornoza:2013ema}.
  The shaded area represents the disfavoured region (at 95\% confidence level) on the
  mass of the LKP from the LHC~\cite{Deutschmann:2017bth}. }
\label{fig:sd_limit}
\end{figure}
\section{Results}
Figure~\ref{fig:data} shows that no statistically significant deviation from the expected background was detected in the IceCube solar analysis, so a 90\% confidence
level limit on the $B^1$-proton cross section can be extracted from the data by setting a limit to $\Gamma_A$. 
The amount of expected signal events per unit of time can be estimated as 
\begin{equation}
  \frac{d\mu_s}{dt}\simeq\frac{\mu_s}{\tau}
\label{eq:mu_rate}
\end{equation}
where $\tau$ is the total exposure time,  532 days in our case. We proceed with a ``counting experiment'', comparing the number of background events to
the number of observed events extracted from Figure~\ref{fig:data}, and construct Poisson confidence intervals for the signal strength $\mu_s$, calculated
with the Neyman method using the algorithm developed in~\cite{Conrad:2002kn}, including the detector systematic uncertainties from~\cite{Aartsen:2016zhm}. 
A limit on $\mu_s$ is easily translated to a limit on $\Gamma_A$ through equation~(\ref{eq:mu_s}) and further to limits on $\sigma^{SI}$ and  $\sigma^{SD}$ through equation~(\ref{eq:Xsection}).  
Table~\ref{tab:results} shows the results for a series of $B^1$ masses. Limits at 90\% confidence level on the signal strength ($\mu_s$), the annihilation rate in the Sun ($\Gamma_A$), the muon flux at the detector above 1~GeV ($\Phi_\mu$) and the spin-independent and spin-dependent cross sections  ($\sigma^{SI}$ and  $\sigma^{SD}$) are shown, along with the median angular resolution ($\Delta\theta_\nu$) and mean muon energy at the detector ($\left<E_\mu\right>$) for each signal model.

\begin{table*}[t]
\centering
\caption{90\% confidence level upper limits on the signal, the annihilation rate in the Sun, the muon flux at the detector above 1~GeV and the spin independent and spin dependent $B^1$-proton cross sections for several values of the $B^{1}$ mass. The last two columns show the median angular resolution and the muon mean energy at the detector.}
\label{tab:results}
\begin{tabular}{cccccccc } 
\hline
$m_{B^1}$ & $\mu_s$ & $\Gamma_A$ & $\Phi_\mu$ & $\sigma^{SI}$ & $\sigma^{SD}$ & $\Delta\theta_\nu$ & $\left<E_\mu\right>$\\
(GeV) & & (ann$\cdot$ s$^{-1}$) & (km$^{-2}$y$^{-2}$) & (cm$^2$) & (cm$^2$) & ($^{\circ}$) & (GeV)\\
\hline
25 & 907.6 & $8.3\cdot10^{24}$ & $2.0\cdot10^4$ & $5.1\cdot10^{-41}$ & $6.0\cdot10^{-39}$ & $\pm18.1$ & $6.8$\\
50 & 186.7 & $1.3\cdot10^{23}$ & $1.1\cdot10^3$ & $1.4\cdot10^{-42}$ & $2.8\cdot10^{-40}$ & $\pm7.5$ & $13.3$\\
75 & 111.8 & $2.0\cdot10^{22}$ & $3.5\cdot10^2$ & $3.4\cdot10^{-43}$ & $9.3\cdot10^{-41}$ & $\pm5.1$ & $19.3$\\
100 & 114.0 & $8.6\cdot10^{21}$ & $2.5\cdot10^2$ & $2.0\cdot10^{-43}$ & $6.7\cdot10^{-41}$ & $\pm5.3$ & $25.8$\\
250 & 80.5 & $8.9\cdot10^{20}$ & $98$ & $7.1\cdot10^{-44}$ & $4.03\cdot10^{-41}$ & $\pm3.6$ & $54.7$\\
500 & 73.1 & $3.5\cdot10^{20}$ & $76$ & $8.3\cdot10^{-44}$ & $6.3\cdot10^{-41}$ & $\pm2.9$ & $88.0$\\
700 & 72.16 & $2.7\cdot10^{20}$ & $73$ & $1.1\cdot10^{-43}$ & $9.4\cdot10^{-41}$ & $\pm2.7$ & $107.6$\\
900 & 71.77 & $2.4\cdot10^{20}$ & $71$ & $1.6\cdot10^{-43}$ & $1.4\cdot10^{-40}$ & $\pm2.6$ & $117.1$\\ 
1100 & 71.5 & $2.2\cdot10^{20}$ & $71$ & $2.1\cdot10^{-43}$ & $1.9\cdot10^{-40}$ & $\pm2.5$ & $126.9$\\ 
1500 & 71.4 & $2.0\cdot10^{20}$ & $71$ & $3.4\cdot10^{-43}$ & $3.2\cdot10^{-40}$ & $\pm2.4$ & $135.1$\\ 
3000 & 71.3 & $1.8\cdot10^{20}$ & $70$ & $1.2\cdot10^{-42}$ & $1.1\cdot10^{-39}$ & $\pm2.4$ & $148.3$\\ 
4000 & 71.2 & $1.8\cdot10^{20}$ & $70$ & $2.0\cdot10^{-42}$ & $2.0\cdot10^{-39}$ & $\pm2.3$ & $150.3$\\ 
5000 & 71.2 & $1.8\cdot10^{20}$ & $71$ & $3.1\cdot10^{-42}$ & $3.1\cdot10^{-39}$ & $\pm2.4$ & $151.5$\\ 
\hline
\end{tabular}
\end{table*}

Figure~\ref{fig:sd_limit}  shows $\sigma^{SD}$ versus $B^1$ mass for the current analysis (blue dots) and the previously published analysis by IceCube in the 22-string configuration (orange dots), where it can be seen that the constraints have been improved by up to one  order of magnitude. The figure also shows the results from the ANTARES collaboration~\cite{Zornoza:2013ema} (black curve). 
The shaded area shows the disfavoured mass region for the first Kaluza-Klein excitation obtained from searches for UED at the LHC~\cite{Deutschmann:2017bth}, where a limit on 1/R (GeV) is obtained by combining several searches for events with large missing transverse momentum or monojets by ATLAS and CMS at 8~TeV and 13~TeV center of mass energies. Collider searches provide a complementary approach to indirect searches for dark matter in the form of Kaluza-Klein modes with neutrino telescopes, being competitive in different regions of the LKP mass range. Additional constraints from cosmology (that the $B^1$ must have a relic density compatible with the estimated dark matter density from CMB measurements) require the mass of the $B^1$ to be below $\sim$1.6 TeV~\cite{Blennow:2009ag,Belanger:2010yx,Arrenberg:2008np}. Thus, taken all results together, the allowed parameter space for the $B^1$ to constitute
the only component of dark matter in the Universe is currently quite restricted, but non-minimal UED models, not probed here, can still provide viable dark matter candidates~\cite{Flacke:2017xsv}.

From the experimental point of view, a few simplifying assumptions have been made to obtain the presented results. Uncertainties on the solar structure model and dark matter velocity distribution have been ignored, and binned data (from Figure~\ref{fig:data}) have been used instead of a continuous sample of individual events, which limits the statistical power of the analysis. 
Even with these approximations, the IceCube limit presented in this letter considerably improves over those published in~\cite{Abbasi:2009vg} and~\cite{Zornoza:2013ema}.


\begin{thebibliography}{}

\bibitem{Garrett:2010hd}
  K.~Garrett and G.~Duda, ``Dark Matter: A Primer,''
  Adv.\ Astron.\  {\bf 2011}, 968283 (2011)
  
\bibitem{Bertone:2004pz} 
  G.~Bertone, D.~Hooper and J.~Silk, ``Particle dark matter: Evidence, candidates and constraints'',
  Phys.\ Rept.\  {\bf 405}, 279 (2005)

\bibitem{Kaluza:1921tu} 
  T.~Kaluza, ``Zum Unittatsproblem der Physik'',
  Sitzungsber.\ Preuss.\ Akad.\ Wiss.\ Berlin (Math.\ Phys.\ ) {\bf 1921}, 966 (1921).\\
  Also Int.\ J.\ Mod.\ Phys.\ D {\bf 27}, no. 14, 1870001 (2018)
  
\bibitem{Klein:1926tv} 
  O.~Klein, ``Quantum Theory and Five-Dimensional Theory of Relativity'',  Z.\ Phys.\  {\bf 37}, 895 (1926)
  
\bibitem{Cheng:2002ej}
  H.~C.~Cheng, J.~L.~Feng and K.~T.~Matchev, ``Kaluza-Klein dark matter'',' Phys.\ Rev.\ Lett.\  {\bf 89}, 211301  (2002).

\bibitem{Hooper:2002gs} 
  D.~Hooper and G.~D.~Kribs, ``Probing Kaluza-Klein dark matter with neutrino telescopes'', Phys.\ Rev.\ D {\bf 67}, 055003 (2003)
  
\bibitem{Servant:2002aq} 
  G.~Servant and T.~M.~P.~Tait, ``Is the lightest Kaluza-Klein particle a viable dark matter candidate?'', Nucl.\ Phys.\ B {\bf 650}, 391 (2003)

\bibitem{Hooper:2007qk} 
  D.~Hooper and S.~Profumo, ``Dark matter and collider phenomenology of universal extra dimensions'', Phys.\ Rept.\  {\bf 453}, 29 (2007)
  
\bibitem{Blennow:2009ag}
  M.~Blennow, H.~Melbeus and T.~Ohlsson, ``Neutrinos from Kaluza-Klein dark matter in the Sun,'' JCAP {\bf 1001}, 018 (2010)
  
\bibitem{Flacke:2017xsv}
  T.~Flacke, D.~W.~Kang, K.~Kong, G.~Mohlabeng and S.~C.~Park,  ``Electroweak Kaluza-Klein Dark Matter'', JHEP {\bf 1704} 041 (2017).

\bibitem{Freitas:2008vh}
  A.~Freitas and U.~Haisch, ``Anti-B $\rightarrow$ X(s) gamma in two universal extra dimensions'',
  Phys.\ Rev.\ D {\bf 77}, 093008 (2008)

\bibitem{Haisch:2007vb}
  U.~Haisch and A.~Weiler, ``Bound on minimal universal extra dimensions from anti-B $\rightarrow$ X(s)gamma,''
  Phys.\ Rev.\ D {\bf 76}, 034014 (2007)
  
\bibitem{Spergel:1984re}
  D.~N.~Spergel and W.~H.~Press, ``Effect of hypothetical, weakly interacting, massive particles on energy transport in the solar interior,''
  {\em Astrophys.\ J.}  {\bf 294}, 663 (1985)
  
\bibitem{Press:1985ug}
  W.~H.~Press and D.~N.~Spergel, ``Capture by the sun of a galactic population of weakly interacting massive particles,'', Astrophys.\ J.\  {\bf 296}, 679, (1985)

\bibitem{Gaisser:1986ha}
  T.~K.~Gaisser, G.~Steigman and S.~Tilav, ``Limits on Cold Dark Matter Candidates from Deep Underground Detectors,''
  {\em Phys.\ Rev.} D {\bf 34} (1986) 2206.

\bibitem{IceCube:2011aj}
  R.~Abbasi {\it et al.},``Multi-year search for dark matter annihilations in the Sun with the AMANDA-II and IceCube detectors,''
  Phys.\ Rev.\ D {\bf 85},042002 (2012)

\bibitem{Aartsen:2016zhm}
  M.~G.~Aartsen {\it et al.},
  ``Search for annihilating dark matter in the Sun with 3 years of IceCube data,''
  Eur.\ Phys.\ J.\ C {\bf 77} no.3,  146,  (2017)
  Erratum: [Eur.\ Phys.\ J.\ C {\bf 79} no.3,  214,  (2019)]

\bibitem{Adrian-Martinez:2016gti}
  S.~Adrian-Martinez {\it et al.}, ``Limits on Dark Matter Annihilation in the Sun using the ANTARES Neutrino Telescope,''
  Phys.\ Lett.\ B {\bf 759}, 69 (2016)

\bibitem{Avrorin:2014swy}
  A.~D.~Avrorin {\it et al.}, ``Search for neutrino emission from relic dark matter in the Sun with the Baikal NT200 detector,''
  Astropart.\ Phys.\  {\bf 62}, 12 (2015) 

\bibitem{Abbasi:2009vg}
  R.~Abbasi {\it et al.},
  ``Limits on a muon flux from Kaluza-Klein dark matter annihilations in the Sun from the IceCube 22-string detector,''
  Phys.\ Rev.\ D {\bf 81}, 057101 (2010)

\bibitem{Zornoza:2013ema}
  J.~D.~Zornoza, ``Search for Dark Matter in the Sun with the ANTARES Neutrino Telescope in the CMSSM and mUED frameworks,''
  Nucl.\ Instrum.\ Meth.\ A {\bf 725}, 76 (2013)
  
\bibitem{Abbasi:2009uz}
  R.~Abbasi {\it et al.},
  ``Limits on a muon flux from neutralino annihilations in the Sun with the IceCube 22-string detector,''
  Phys.\ Rev.\ Lett.\  {\bf 102}, 201302,  (2009)

\bibitem{Belanger:2010yx}
  G.~Belanger, M.~Kakizaki and A.~Pukhov, ``Dark matter in UED: The Role of the second KK level,''  JCAP {\bf 1102}, 009 (2011)

\bibitem{Wikstrom:2009kw}
  G.~Wikstr\"om and J.~Edsj\"o,
  ``Limits on the WIMP-nucleon scattering cross-section from neutrino telescopes,''
  JCAP {\bf 0904}, 009 (2009)

\bibitem{Edsjo:XXX} J. Edsj\"o, J. Elevant and C. Niblaeus, WimpSim Neutrino Monte Carlo, {\texttt{http://wimpsim.astroparticle.se}}

\bibitem{Serenelli:2009yc}
  A.~Serenelli, S.~Basu, J.~W.~Ferguson and M.~Asplund, ``New Solar Composition: The Problem With Solar Models Revisited,''
  Astrophys.\ J.\  {\bf 705}, L123 (2009)
  
\bibitem{Bringmann:2018lay}
  T.~Bringmann, J.~Edsj\"o, P.~Gondolo, P.~Ullio and L.~Bergstr\"om,
  ``DarkSUSY 6: An Advanced Tool to Compute Dark Matter Properties Numerically,''
  JCAP {\bf 1807}, 033, (2018). Also {\texttt{http://www.darksusy.org}}

\bibitem{Conrad:2002kn}
  J.~Conrad, O.~Botner, A.~Hallgren and C.~P\'erez de los Heros, 
  ``Including systematic uncertainties in confidence interval construction for Poisson statistics,''
  Phys.\ Rev.\ D {\bf 67}, 012002 (2003)

\bibitem{Deutschmann:2017bth}
  N.~Deutschmann, T.~Flacke and J.~S.~Kim, ``Current LHC Constraints on Minimal Universal Extra Dimensions,''
  Phys.\ Lett.\ B {\bf 771}, 515 (2017)

\bibitem{Beuria:2017jez}
  J.~Beuria, A.~Datta, D.~Debnath and K.~T.~Matchev, ``LHC Collider Phenomenology of Minimal Universal Extra Dimensions,''
  Comput.\ Phys.\ Commun.\  {\bf 226}, 187 (2018)
  
\bibitem{Arrenberg:2008np}
  S.~Arrenberg, L.~Baudis, K.~Kong, K.~T.~Matchev and J.~Yoo, ``Kaluza-Klein Dark Matter: Direct Detection vis-a-vis LHC,''
  PoS IDM {\bf 2008}, 059 (2008)

\end{thebibliography}
\end{document}